# Comments on "Partition function of improved Tietz oscillators, C.S. Jia et all, Chemical Physics Letters 676, 2017, 150"

I. H. Umirzakov

Institute of Thermophysics, Lavrentev prospect, 1, 630090 Novosibirsk, Russia

e-mail: cluster125@gmail.com

**Abstract** The closed-form expression for the quantum partition function of the improved Tietz oscillator is obtained using the Voronoi summation formula.

**1.** The aim of [1] was to obtain closed-form expression for the quantum vibrational partition function for the improved Tietz potential energy model [2]. According to [1] the terms with $m \neq 0$ include the quantum corrections in following equation

$$\sum_{n=0}^{N} f(n) = \frac{1}{2}[f(0) - f(N+1)] + \sum_{m=-\infty}^{\infty} \int_{0}^{N+1} f(x) \exp(-i 2\pi m x) dx, \qquad (1)$$

where $i$ is the imaginary unit. Neglecting the quantum corrections Eq. 1 was represented in [1] as

$$\sum_{n=0}^{N} f(n) = \frac{1}{2}[f(0) - f(N+1)] + \int_{0}^{N+1} f(x) dx. \qquad (2)$$

But neglecting the quantum corrections contradicts to the aim of [1] (Eqs. 1-2 are equivalent to Eqs. 9-10 [1]).

Eq. 11 [1] for the partition function was obtained using Eq. 2. As one can see Eq. 11 [1] takes into account the vibrational quantum states. This is the second contradiction in [1].

According to [1] Eq. 1 is the Poisson summation formula (PSF). PSF does not consider the quantum corrections [4]. Therefore the terms with $m \neq 0$ in Eq. 1 does not include the quantum corrections. This is the third contradiction in [1].

There is no a proof in [1] that the terms with $m \neq 0$ in Eq. 1 include the quantum corrections to the classical partition function.

So, there is the inner inconsistency in [1].

**2.** The Poisson summation formula is valid if $\int_{-\infty}^{\infty} |f(x)| dx < \infty$ [4]. One can see that

$\int_{-\infty}^{\infty} |f(v)| dv = \infty$ for $f(v) = \exp(-E_v / kT)$, where $k$ is the Boltzmann constant, $T$ is the temperature and the energy $E_v$ of $v$-th vibrational quantum level is defined from Eq. 2 [1]. Therefore Eqs. 1-2 cannot be used to calculate the partition function $Q = \sum_{v=0}^{N} \exp(-E_v / kT)$, where $N$ is the number of exited bonded states.

**3.** Eq. 2 can be represented as

$$\int_{0}^{N+1} f(x) dx = \sum_{n=0}^{N} \frac{1}{2}[f(n) + f(n+1)]. \qquad (3)$$

It is easy to establish that Eq. 3 is equivalent to

$$\int_0^a f(x)dx = \sum_{n=0}^{a/\Delta-1} \frac{1}{2}[f(n\Delta) + f[(n+1)\Delta]], \qquad (4)$$

where $\Delta = 1$ and $a = (N+1)\Delta$. One can see that Eq. 4 is correct if $f(x)$ is a linear function or it is equal to a constant [3].

According to the trapezoid rule [3]

$$\int_0^a f(x)dx = \sum_{n=0}^{a/\Delta-1} 0.5 \cdot [f(n \cdot \Delta) + f[(n+1)\Delta]]\bigg|_{\Delta \to 0}. \qquad (5)$$

We conclude from the comparison of Eqs. 4 and 5 that Eq. 4 is incorrect for a non-linear function $f(x)$. Hence Eqs. 2-3 are incorrect for a non-linear function $f(x)$.

One can conclude from above comments **2** and **3** that Eqs. 11 [1] and the closed-form expressions (Eqs. 14-16 [1]), obtained in [1] using Eq. 2 for the partition function, are incorrect. Eqs. 14-16 [1] are main results of [1].

**4**. The function $f(v) = \exp(-E_v/kT)$ is the non-linear function of $v$. Eq. 2 is incorrect for a non-linear function $f(v)$. Therefore Eqs. 11 and 14-16 [1] are incorrect.

**5**. Fig. 2 [1] confirms the incorrectness of Eq. 14 [1] for the partition function of $CO$ molecule.

**6**. One can conclude on the basis of above comments **1-5** that the conclusions [1] based on the incorrect Eqs. 14-16 [1] could be incorrect.

There exists the exact Voronoi summation formula

$$\sum_{n=0}^{N} f(n) = \int_0^N f(x)dx + 2\sum_{m=1}^{\infty} \int_0^N f(x)\cos(2\pi mx)dx \qquad (6)$$

for smooth function $f(x)$ of bonded variation [5-7] which cannot obey the condition $\int_{-\infty}^{\infty} |f(x)|dx < \infty$. Eq. 6 can be represented as

$$\sum_{n=0}^{N} f(n) = \sum_{m=-\infty}^{\infty} \int_0^N f(x)\exp(-i2\pi mx)dx. \qquad (7)$$

Comparing Eqs. 1 and 7 one can conclude that Eq. 1 is incorrect.

If we neglect in Eq. 6 the terms with $m \neq 0$ as it was done in [1] then using $\sum_{n=0}^{N} f(n) = \int_0^N f(x)dx$ and repeating the mathematical operations done in [1] we obtain

$$Q = \int_0^{v_{max}} e^{-\beta\left(D_e - \lambda\left(\frac{\delta_1}{x+\delta_2} - \frac{x+\delta_2}{2}\right)^2\right)} dx$$

instead of the incorrect Eq. 11 [1]. One can easily see that:
- the upper limit $v_{max} + 1$ must be replaced by $v_{max}$ in the integral in left hand side of Eq. 12 [1];
- it is necessary to replace $v_{max}$ by $v_{max} - 1$ in Eqs. 18, 20 and 21 [1];

- the term $e^{\beta\lambda b_1^2} - e^{\beta\lambda b_2^2}$ in the square brackets in the right hand sides of Eqs. 14 and 16 [1] must be removed, and therefore Eq. 15 [1], which is the power series of Eq. 14 for $\beta \ll 1$, is incorrect;

- it is necessary to use the correct relation $b_2 = \dfrac{\delta_1}{v_{max} + \delta_2} - \dfrac{v_{max} + \delta_2}{2}$ instead of the incorrect one $b_2 = \dfrac{\delta_1}{v_{max} + 1 + \delta_2} - \dfrac{v_{max} + 1 + \delta_2}{2}$ in Eqs. 12, 14 and 15 [1];

- it is necessary to use the correct relation $c_v = e^{a v_{max}^2 - b v_{max}}$ in Eq. 21 [1] instead of incorrect one $c_v = e^{a(v_{max}+1)^2 - b(v_{max}+1)}$.